\documentclass[journal]{rmaa}

\title {The Red Stellar Contents of Three Selected Fields of
The Local Group Dwarf Irregular Galaxy IC\,1613.}

\author{J.~Borissova\altaffilmark{1}, L.~Georgiev\altaffilmark{2}, R.~Kurtev\altaffilmark{3}, M.~Rosado\altaffilmark{2}, \\
V.~D.~Ivanov\altaffilmark{4}, M.~Richer\altaffilmark{2} and M.~Valdez-Guti\'errez\altaffilmark{5}
}

\altaffiltext{1}{Institute of Astronomy, Bulgarian Academy of Sciences, Sofia, Bulgaria}
\altaffiltext{2}{Instituto de Astronom\'{\i}a, Universidad Nacional Aut\'onoma de M\'exico, M\'exico}
\altaffiltext{3}{Department of Astronomy, Sofia University, Sofia, Bulgaria}
\altaffiltext{4}{Steward Observatory, The University of Arizona, Tucson, USA}
\altaffiltext{5}{Instituto Nacional de Astrof\'\i sica, Optica y Electr\'onica, M\'exico}

\fulladdresses{
\item Jordanka Borissova: Institute of Astronomy, Bulgarian Academy of Sciences,
72~Tsarigradsko chauss\`ee, BG\,--\,1784 Sofia, Bulgaria (jura@haemimont.bg)
\item Leonid Georgiev, Margarita Rosado and Michael Richer: Instituto de Astronom\'{\i}a, Universidad Nacional Aut\'onoma de M\'exico, Apartado postal 70-264, 04510 M\'exico, D.F., M\'exico (georgiev,margarit,richer@astroscu.unam.mx)
\item Radostin Kurtev: Department of Astronomy, Faculty of Physics, Sofia University, BG\,--\,1164 Sofia, Bulgaria (kurtev@phys.uni-sofia.bg)
\item Valentin Ivanov: Steward Observatory, The University of Arizona, 933 N. Cherry
Ave, Tucson, AZ 85721 (vdivanov@as.arizona.edu)
\item Margarita Valdez-Guti\'errez: Instituto Nacional de Astrof\'\i sica, Optica y Electr\'onica, Apartado postal 51 y 216, 72000 Puebla, M\'exico (mago@inaoep.mx)
}

\shortauthor{Borissova et al.}
\shorttitle{The Red Stellar Contents of IC\,1613}

\SetVolume{00} \SetFirstPage{00} \SetYear{2000}
\ReceivedDate{1999 May 26} 
\AcceptedDate{1999 May 26} 

\resumen{Este es el resumen, lo cual debe de ser una traducci\'on mas
  o menos fiel de lo que est\'a escrito en el abstract. }

\abstract{We present a moderately-deep $JK$ photometry for
three selected areas of the dwarf irregular galaxy IC\,1613. The
color-magnitude diagrams contain a mixture of red supergiants,
asymptotic giant branch stars and the brightest red giant stars.
The  red supergiants are massive (20 - 25$\cal M_\odot$)  and  young
--- with ages between  8 and 25 Myr.  The most important result is
the evidence of the decreasing density of the intermediate age AGB population in
the vicinity of the HII regions in the galaxy. We also find age
differences between AGB stars in the main body of the galaxy and
those near the HII regions in the North-East.  The former span a
range in ages between 1 and 10 Gyr, while the latter are younger
than 1 Gyr.  Using  the period-luminosity
relation derived by Madore \& Freedman (1991) and $JK$ magnitudes
of the Cepheid variable V20, we calculated $(m-M)_{K} =
24.37\pm0.2$. The recently discovered Nova (King et al. 1999) was identified
in Field III. Its presence of our images and its brightness 
questioned its classification as a nova.}


\begin{document}

\maketitle

\section{Introduction}
Studies of bright stars in nearby galaxies provide a direct link
to their history of star formation.  Photometric investigations of
red supergiants (RSG) and  asymptotic giant branch (AGB) are
of particular interest, as these stars probe the evolution during
early and intermediate epochs.

Carrying out such studies at near-infrared (NIR) wavelengths have
a number of advantages.  NIR observations are less sensitive to
both the reddening towards and within galaxies.  They allow us to
overcome the effects of line blanketing, which can affect the
spectral energy distributions of moderately metal-rich giants at
optical wavelengths.  In addition, they are the best choice for
determining bolometric corrections and for the analysis of the
bolometric luminosity function of the luminous, cool evolved
stars.  Several recent studies of the red stellar content in
nearby spiral and irregular galaxies exploiting these advantages
include Davidge (1999: M\,33), Davidge (1998: NGC\,55, NGC\,300,
and NGC\,7793), and Cioni et al. (1999: LMC).

Here, we present NIR observations of IC\,1613. IC\,1613 is a
faint, irregular galaxy within the Local Group. Recently, Cole et
al. (1999) studied the history of star formation in the central
part of the galaxy using HST $VI$ photometry.  They found a
dominant old stellar population --- red giant branch and red clump
stars, and a prominent sequence of intermediate age AGB stars.
Cole et al. (1999) deliberately chose their region for study to
avoid known HII regions.  In this paper, we present $JK$
photometry of the three selected areas of IC\,1613.  We chose our
fields to extend the area covered by Cole et al. (1999) towards
the active HII areas. The goal of this work is to analyze the
bright, red stellar content and to determine the distance of
IC\,1613.

\section{Observations and data reduction}

We acquired our data with the \lq\lq CAMILA" infrared camera
attached to the 2.1-m telescope of the Observatorio
Astronom\'{\i}co National at "San Pedro Martir", M\'exico
(Cruz-Gonz\'alez et al. 1996).  CAMILA is based upon a NICMOS3
($256\times 256$ pixels) detector.  The scale was $0.85$
arcsec/pixel, which resulted in a field size of about $3.6 \arcmin
\times 3.6 \arcmin$. A set of $JK$ frames was taken on October
13-15, 1998. The  seeing during these observations was between $1
- 1.2 \arcsec$ with stable and very good photometric conditions.
Ten UKIRT (Casali \& Hawarden  1992)  standard stars were observed
during the observations. All three fields were obtained with equal
exposure times of 960 sec.

\begin{figure}
\begin{center}
\includegraphics[width=\columnwidth]{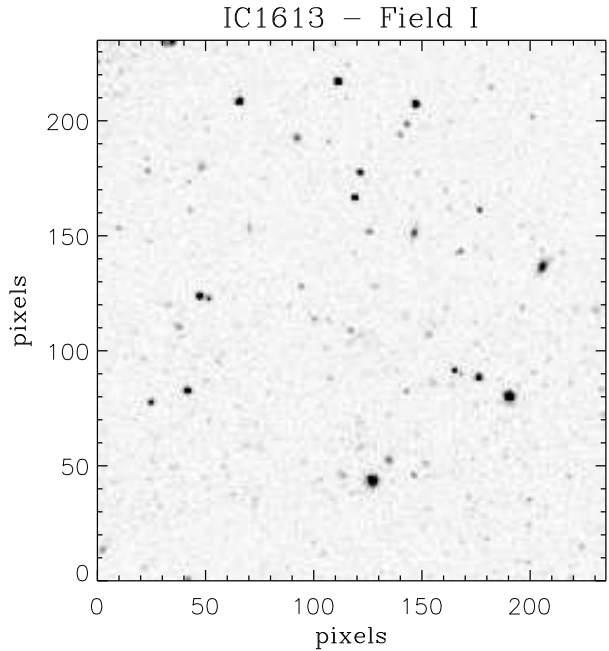}
\end{center}  
\caption{Final $J$  image of IC\,1613 Field I. North is up and East is to the left.
  \label{Fig1}}
\end{figure}

\begin{figure}
\begin{center}
\includegraphics[width=\columnwidth]{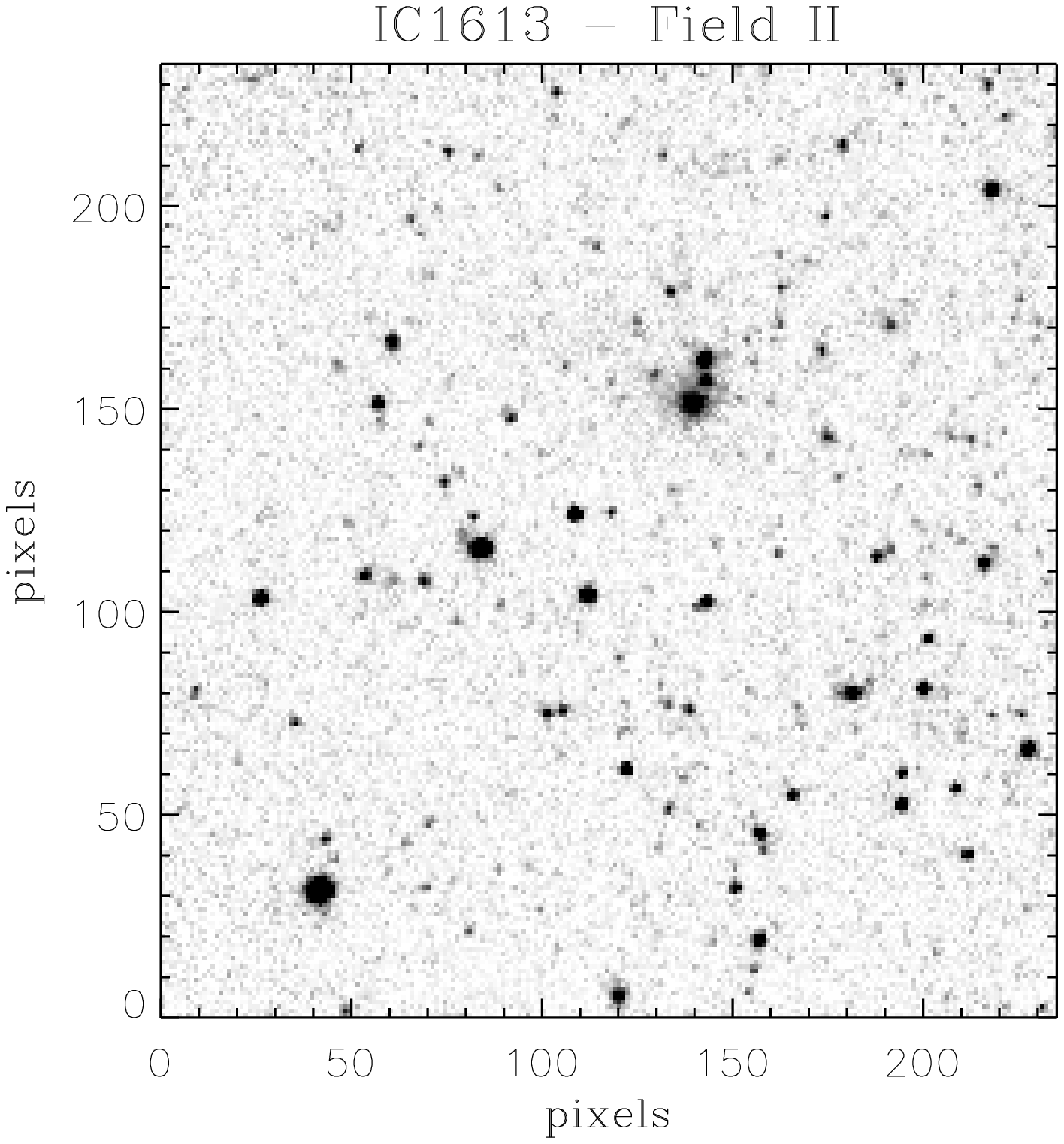}
\end{center}  
\caption{ Final $J$  image of IC\,1613 Field II. North is up and East is to the left.
  \label{Fig2}}
\end{figure}

\begin{figure}
\begin{center}
\includegraphics[width=\columnwidth]{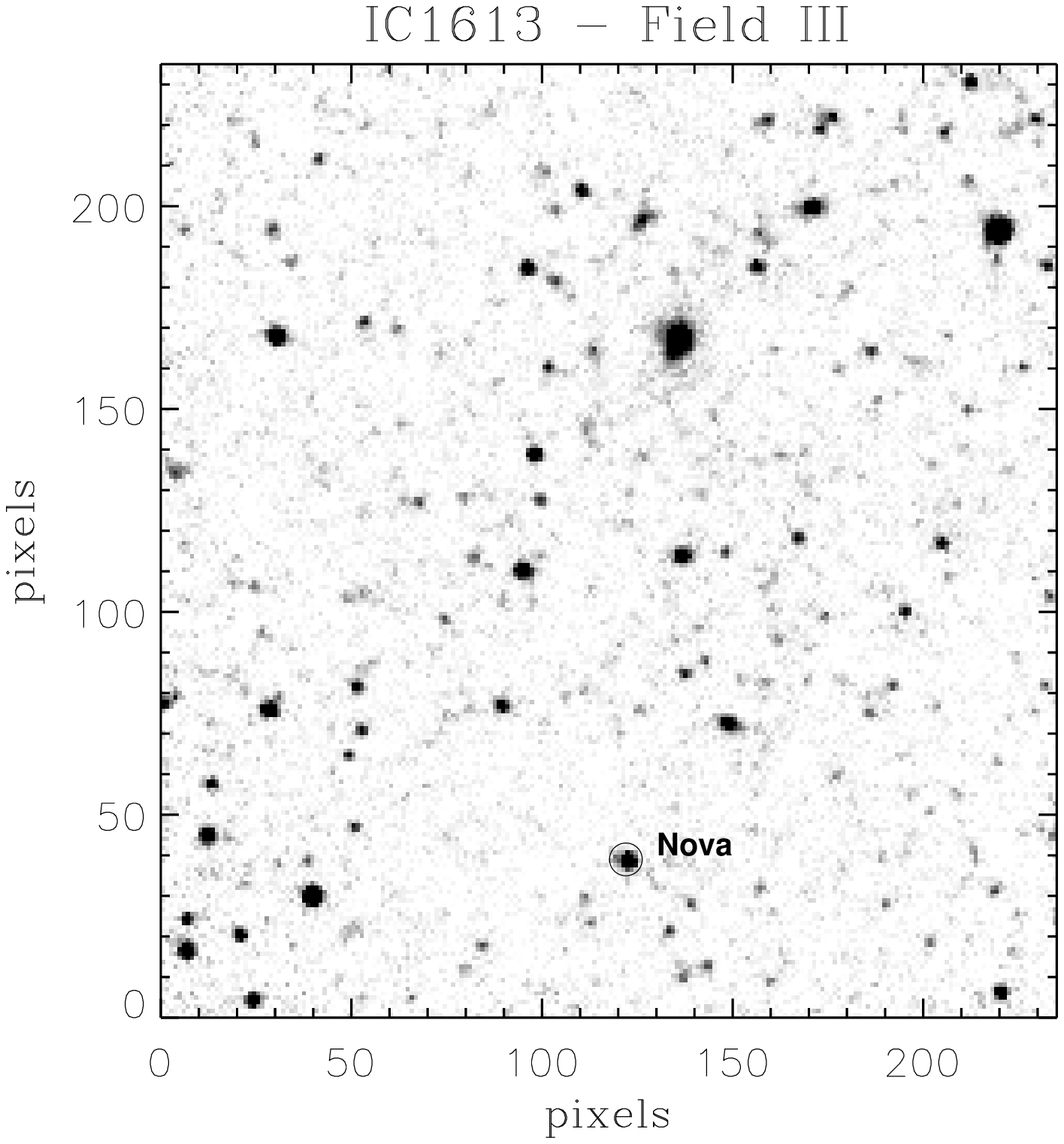}
\end{center}  
\caption{Final $J$  image of IC\,1613 Field III. North is up and East is to the left.
  \label{Fig3}}
\end{figure}

\begin{figure}
\begin{center}
\includegraphics[width=\columnwidth]{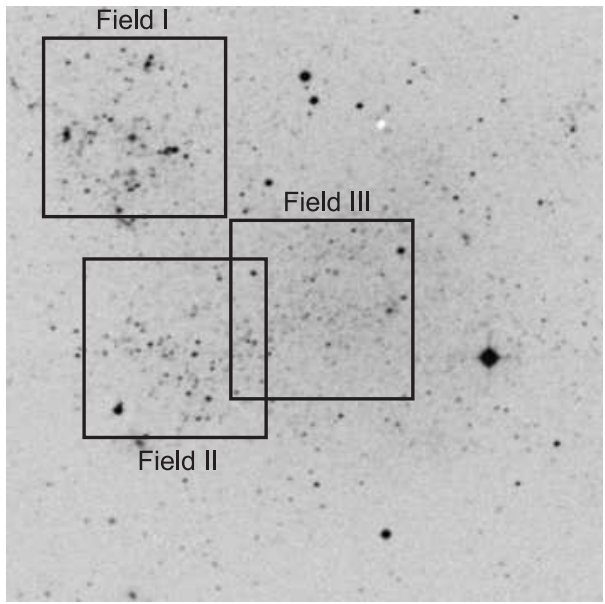}
\end{center}  
\caption{The Digitized Sky Survey image of IC\,1613 (North is up and East is to
the left). Our three $3.6\arcmin \times 3.6 \arcmin$ fields are
indicated.
  \label{Fig4}}
\end{figure}

The IRAF data reduction package was used to carry out the basic
image reduction. The final median-combined $J$ images of the
fields are shown in Fig.~\ref{Fig1}, Fig.~\ref{Fig2} and
Fig.~\ref{Fig3}.  The three fields are indicated on a print of a
$B$ plate from STScl Digitized Sky Survey  (Fig.~\ref{Fig4}).
Field I is centered on the HII regions in the north-east quadrant
of the galaxy where the most active ongoing star formation is
occurring (Sandage 1971).  Fields II and III were chosen to be
within of the main body of IC\,1613.  There are six associations
Hodge (1978) in Field I --- A10, A11, A12 A13, A14 and A17, Field
II include Hodge (1978) associations A8 and A9 and Field III
include A5, A6 and A7.  

The stellar photometry of the median-combined frames was performed with the point$-$spread function (PSF) fitting routine ALLSTAR available in DAOPHOT (Stetson 1993). 
The ALLSTAR magnitudes were calibrated against aperture photometry performed on selected stars which had neighbors subtracted. We then rejected all stars which had values of $\chi^2\ > \ 2$ and the stars with formal errors from the PSF fitting greater than 0.25 in all frames. The instrumental values were transformed to the standard $J$ and $K$ system by means of the equations:

\begin{eqnarray*}
J  &=& -4.447_{(\pm0.09)} +1.002_{(\pm0.05)}\,j - \\
&&- 0.123_{(\pm0.03)}\,(j-k)-0.168_{(\pm0.06)}\,X
\end{eqnarray*} 

\begin{eqnarray*}
K  &=& -4.724_{(\pm0.08)} +0.992_{(\pm0.03)}\,k - \\
&&-0.173_{(\pm0.05)}\,(j-k)-0.154_{(\pm0.03)}\,X
\end{eqnarray*}
where $j$ and $k$ are the instrumental magnitudes, $X$ is the airmass and ($j-k$) is the color of the object. The errors are the statistical  errors in the coefficients. The zero-point errors of the transformation are  0.02 in $J$ and 0.05 in $K$. 

The artificial star simulations necessary for the completeness correction are used to determine the measurement errors. The $K$ band "input-output" mean differences between "added" and "recovered" artificial stars as a function of specified magnitude interval for Field I, Field II and Field III are given  in Table~1.  In column 5 and 6, 10 and 11, 15 and 16 are given mean differences between actual and measured magnitudes of synthetic stars recovered in specified interval of one magnitude and the error of the mean difference. The mean differences and errors in $J$ band are within the same intervals . In summary, the median scatter is smaller than 0.1 mag down to 17.5 mag and increases up to 0.2 in the faintest bins.

\begin{table*}\tabcolsep=3pt\small
\caption{Completeness and mean errors from artificial stars in $K$ band}
\begin{center}
\begin{tabular}{cccccccccccccccc}
\hline
Range &\multicolumn{5}{c}{Field\,I}&\multicolumn{5}{c}{Field\,II}& \multicolumn{5}{c}{Field\,III}\\
$K$ &$K_{\rm input}$ &$K_{\rm output}$&$r$&$\Delta{M}$&$\sigma$&$K_{\rm input}$ &$K_{\rm output}$&$r$&$\Delta{M}$&$\sigma$ &$K_{\rm input}$ &$K_{\rm output}$&$r$&$\Delta{M}$&$\sigma$ \\
\hline
13-14&135&134&0.99&0.02&0.03&126&123&0.98&-0.03&0.04&161&158&0.98&0.02&0.02\\
14-15&147&145&0.99&0.03&0.02&136&133&0.98&-0.01&0.03&169&170&1.00&-0.02&0.04\\	
15-16&129&128&0.99&0.0&0.05&128&126&0.98&0.05&0.04&146&142&0.97&0.05&0.06\\	 
16-17&120&116&0.97&0.08&0.10&113&104&0.92&0.06&0.03&143&131&0.92&0.07&0.05\\	
17-18&98&83&0.85&0.13&0.15&121&97&0.80&0.07&0.11&148&124&0.84&0.08&0.15\\
18-19&71&38&0.53&0.18&0.21&126&56&0.44&0.18&0.15&153&76&0.50&0.15&0.19\\
\hline
\end{tabular}
\end{center}
\end{table*}

\section{Completeness and field stars contamination}

The artificial star technique (Aparicio \& Gallart 1995) was used to
investigate the crowding effects. Following step by step the procedure described in 
Aparicio \& Gallart (1995) we generated the tables in $J$ and $K$ bands with artificial stars, which have realistic magnitudes and colors. Such prepared tables were used to generate the synthetic frames for the three investigated fields. They were processed in the same way as original frames. The process was repeated many times and in total 700 stars are analyzed for Filed I,
750 for Field II and 920 for Field III. In Columns 2, 3 and 4 (respectively 7,8 and 9 and 12,13 and 14 for Fields II and III) of the Table~1  are given total number of "input" and "output" stars and the ratio between them in $K$ band.  In summary, the analysis of Table~1 show, that the degree of completeness  is close to 100 per cent down to 16 mag in $K$ and can seriously affect our analysis for magnitudes fainted than 18 mag. As can be seen there is no significant differences in the completeness factors between the three investigated fields. 

\begin{figure}
\begin{center}
\includegraphics[width=\columnwidth]{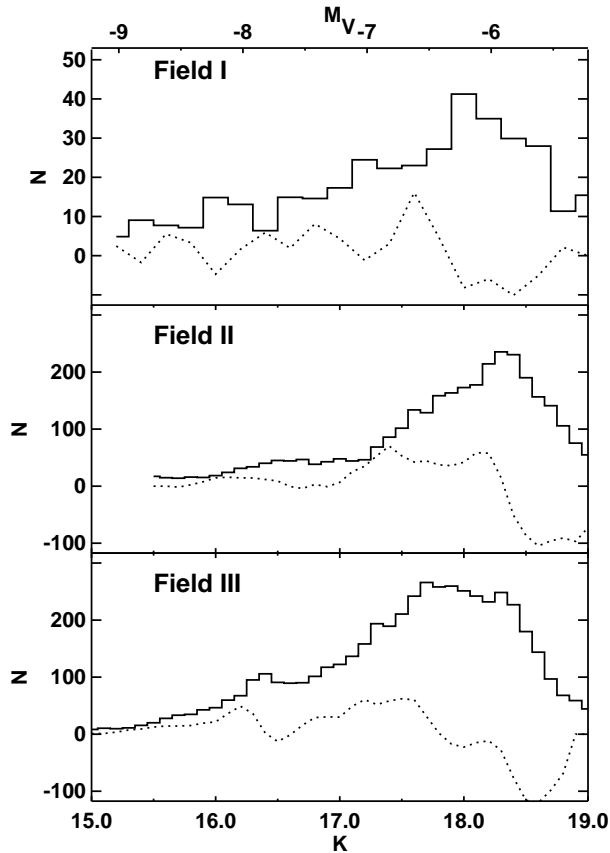}
\end{center}  
\caption{The differential luminosity function for all stars in Field I,
Field II and Field III. \lq\lq N" is the number of stars per
magnitude interval. The dotted lines are the output from the Sobel
edge-detection filters (see Section~\ref{Sect6} for
details).
  \label{Fig5}}
\end{figure}

The luminosity functions (LF) for the three fields are shown in Fig.~\ref{Fig5}.
The decrease in counts at the faint end also indicates that the incompleteness  becomes
significant when $K = 18.0-18.2$.

Field stars should not seriously affect the structure of the
color-magnitude diagram because IC\,1613 is situated far from the
galactic plane ($b=-61^\circ ,\ l=131^\circ$). The  field star
densities predicted by the Bahcall \& Soneira (1980) model imply
that the contamination from our Galaxy is negligible.  We applied
no correction for contamination by field star.

\section{ Color-magnitude diagrams}

\begin{figure}
\begin{center}
\includegraphics[width=\columnwidth]{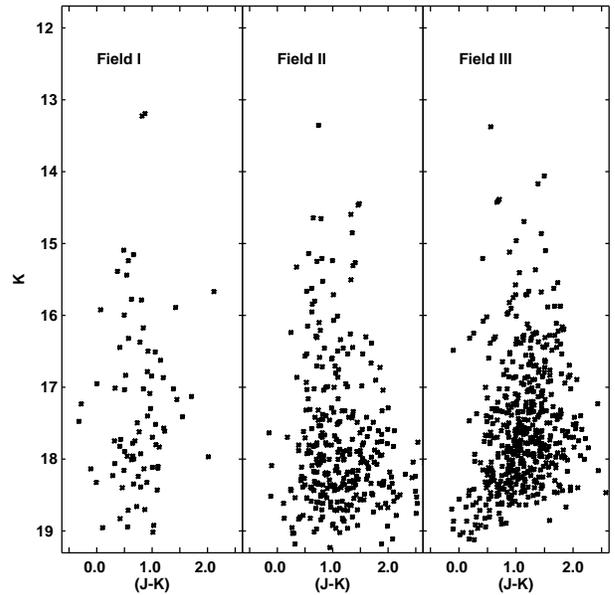}
\end{center}  
\caption{$(J-K,K)$ color-magnitude diagrams obtained for IC\,1613.  \label{Fig6}}
\end{figure}

Fig.~\ref{Fig6} shows the $(J-K,K)$ color-magnitude diagrams (CMD)
of our three fields in IC\,1613. Our final list contains 80 stars
in Field I,  343 stars  in Field II, and 505  stars  in Field III.

The CMDs contain a mixture of red supergiants (RSGs), AGB stars,
and the brightest red giant stars (RGB). Each field contains one
or two stars with $K\approx 13 - 13.5$ and $J-K \approx 1$, which
are the very luminous RSGs, and delineate the peak of the RSG
luminosity function. Close binaries whose components have similar
magnitudes and colors will look like one star of the same color
though brighter in magnitude. The brightest stars were inspected
for blending effects on $VI$ frames with much better scale - 
$0.33\arcsec\,{\rm pixel}^{-1}$ and seeing of $1 \arcsec$. The  $VI$ frames were 
obtained on the 2-m Ritchey-Chretien
telescope of the Bulgarian National Astronomical Observatory
on September 15, 1999 with a Photometrics $1024\times 1024$ camera. 
No such occurrences were found. The presence of a population of bright
RSGs is an indicator of recent star formation in all three fields
in IC\,1613.

In Field I we can see some deficit of stars redder than $(J-K) \approx
1.0$. The analysis of crowding effects (see Table~1) shows that there is no
larger crowding in this field. 
Since Field I is centered on the HII regions of the galaxy the higher internal 
reddening can affect the stars.  The "mean" reddening of the area however determined  
by Georgiev et al. (1999) is $E(B-V)=0.06\pm0.02$. Assuming $E(B-V)=0.03\pm0.02$ 
for Fields II and III (Cole et al. 1999) the resulting difference due to reddening is $0.02$
for $J-K$ color -  the value much smaller than the observing effects. 
The $(B-V,V)$ photometry by Georgiev et al. (1999) centered on the same area as Field I 
can provide independent check of the observed number of the red stars in the infrared. 
The number of stars with $B-V > 0.8$ and $V$ brighter than 21.5 is 92, which is in good 
agreement with the total number of corrected for completeness stars, obtained in Field I 
- 95.7. Of course, deeper photometry is necessary to confirm the observed in Field I 
deficit of stars redder than $J-K = 1.0$.

The stars between $K = 14.5 $ and $K = 15.5 $ and $ 0.5 < (J-K) <
1.5$ are also RSGs with lower brightness. When $K > 15.5 $ both
RSGs and AGBs stars are present, although  there is some color
separation between them --- AGB stars are generally redder than
$(J-K) \approx 1.0$.  At the faint limit of the CMDs, but still
within the completeness limit of the photometry, the brighter red
giants can be seen mixed among with the faintest AGB stars.

\section {Age and Luminosity Function}\label{Sect6}

\begin{figure}
\begin{center}
\includegraphics[width=\columnwidth]{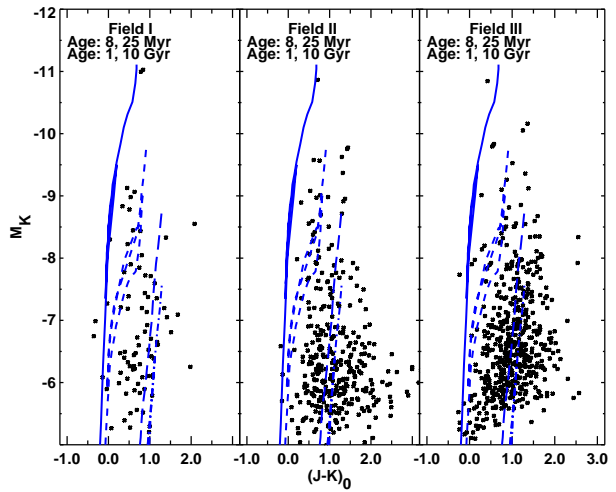}
\end{center}  
\caption{The ($M_K,(J-K)_{0}$) color-magnitude diagrams of IC\,1613 with
isochrones from Padua's library for $8\,10^6, 2.5\,10^7,
10^9$ and $10^{10}$ years superimposed (vertical
lines). 
  \label{Fig7}}
\end{figure}

To convert our observed color-magnitude diagrams to absolute
magnitude and true color, we assumed a foreground reddening, a
distance modulus and a metallicity of $E(B-V)=0.06$ ( Georgiev et
al. 1999), $(m-M) = 24.20\pm0.1$ (Freedman 1988, Saha  et al.
1992, Cole et al. 1999), and $Z=0.004$ (Mateo 1998).  The
resulting color-magnitude diagrams are shown in Fig.~\ref{Fig7}.
The isochrones for $8\,10^6, 2.5\,10^7, 10^9$
and $10^{10}$yr, and for $Z=0.004$ from Padua's library
(see Bertelli et al. 1994) are superimposed on the CMDs
(Fig.~\ref{Fig7}).

The isochrones superimposed upon Fig.~\ref{Fig7} indicate that the
brightest red supergiants in the three fields are very young
--- about 8 Myr. Georgiev et al. (1999) obtained  5 Myr for
brightest red supergiants V32 and V38 in Field I, which is in good
agreement. Cole at al. (1999) also found that in the center of
IC\,1613 the red supergiants are younger than a \lq\lq few
$10^7$yr". Fainter RSGs in IC\,1613 have  $M_{K} <-10$ and are
matched very well  by a 25 Myr isochrone.  

\begin{figure}
\begin{center}
\includegraphics[width=\columnwidth]{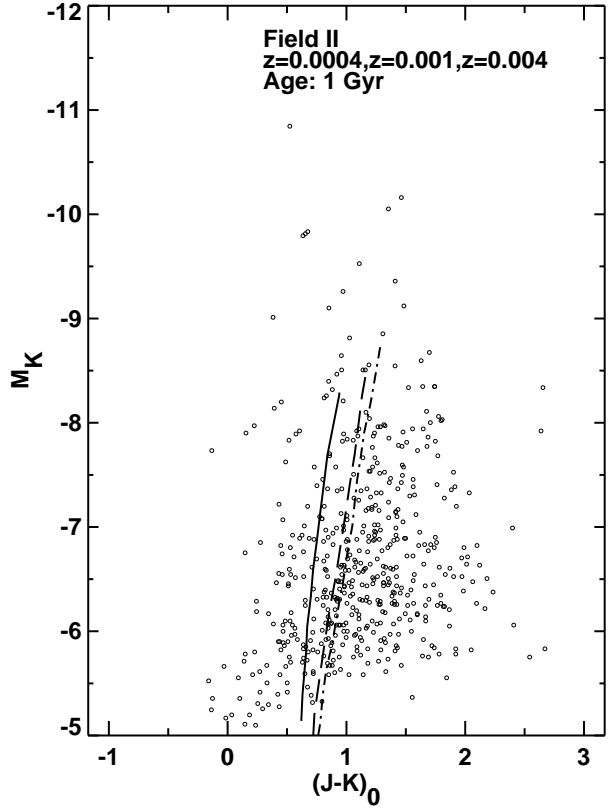}
\end{center}  
\caption{The ($M_K,(J-K)_{0}$) color-magnitude diagrams of IC\,1613 with
isochrones from Padua's library for $10^9$ 
years superimposed (vertical lines) and $Z=0.0004, Z=0.001$ and $Z=0.004$.  \label{Fig8}}
\end{figure}

Most of the AGB stars in Fields II and III are several Gyr old. In Field I most of them
are between the second and third isochrones, and thus are younger
than 1 Gyr.  As well known ( see for example Aparicio, 1999)  the 
colors and life-times of AGB stars depend on the mass loss, bolometric correction 
and metallicity, which are difficult to parameterize. Without information of main sequence 
we will consider above determined range of AGB ages as a rough estimates.   

To investigate the influence of the various metallicity on the AGB ages 
we overplot in Fig.~\ref{Fig8} the isochrones for $10^9$yr
and for $Z=0.0004$, $Z=0.001$ and $Z=0.004$. Indeed, the metal
poorer and metal richer AGB stars can present, 
but it is not possible to separate the metallicity
from age effects using $(J-K,K)$ color-magnitude only. 

\begin{figure}
\begin{center}
\includegraphics[width=\columnwidth]{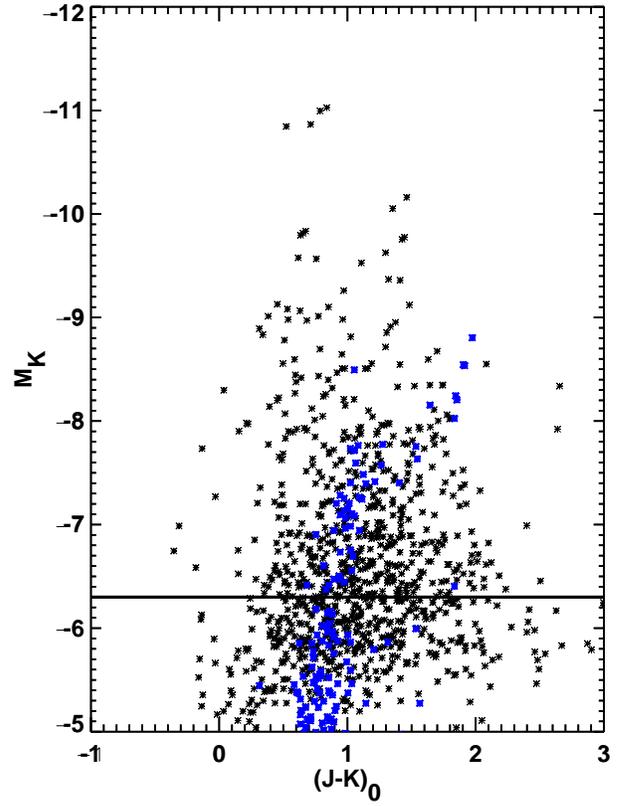}
\end{center}  
\caption{The ($M_K,(J-K)_{0}$) color-magnitude diagrams of all three fields of IC\,1613 (light circles).  $JK$ 
observations (dark circles) of 12 globular clusters
in the Large Magellanic Cloud (Ferraro et al. 1995) are also
drawn.  The solid horizontal line represents the separation
between RGB and AGB stars for the globular clusters.
  \label{Fig9}}
\end{figure}

To determine the tip of AGB stars we used the $K$-band LFs.
Looking at the differential LFs in Fig.~\ref{Fig5} we can see the
AGB tips at $K$ between 16.0 -16.5 in the three fields. To verify
the reality of these edges we convolved the data with a Sobel
edge-detection filter with width of the filter 0.2 (Lee et al. 1993, Sakai at al. 1996). The
filter function outputs are shown as dashed lines in
Fig.~\ref{Fig5}. The limited number of stars in Field I prevents us
from drawing any conclusions. For Fields II and III however, there
are peaks at $K=16.15$ and $K=16.25$, respectively, which mark the
tip of AGB.  Given the reddening and distance modulus adopted
above, this corresponds to approximately $M_K = -8$. 
This value is in agreement with the tip of
the AGB found by interpolating between the isochrones in
Fig.~\ref{Fig7} for ages of 4-5 Gyr, confirming the reality of the
AGB peaks in for Fields II and III. 

The derived power-law exponent for the composite (Fields II and III) LF for magnitude
interval $15.5<K<18$ is 0.51$\pm0.04$. Cole et al. (1999)
found the slope of V-band luminosity function down to $V=24.5$ to
be $0.48\pm0.09$, which is in acceptable agreement.

In Fig.~\ref{Fig9}, the stars from all three fields in IC\,1613 are shown with overploted $JK$ observations of 12 globular clusters in
the Large Magellanic Clouds (Ferraro et al. 1995) (dark circles). The horizontal line is located at the separation
threshold between AGB and RGB stars, as given by Ferraro et al.
(1995). The globular clusters are a good match to our data,
indicating that some long period variables should be located
between $-8 <M_{K} <-9$.  Taking into account Ferraro's (1995)
separation threshold between AGB and RGB stars, the tip of the RGB
should be located around $M_{K}=-6$. The peaks
at $K=18.15$ of Sobel edge-detection output in LFs of Fields II and III
(Fig.~\ref{Fig5}) provide some support of this determination, but
also can be due to incompleteness at fainter magnitudes.
Unfortunately, the RGB stars are at the limit of our photometry
and deeper photometry is necessary to investigate them in more
detail.

\section {Theoretical Hertzsprung--Russell diagrams and Variable stars}

\begin{figure}
\begin{center}
\includegraphics[width=\columnwidth]{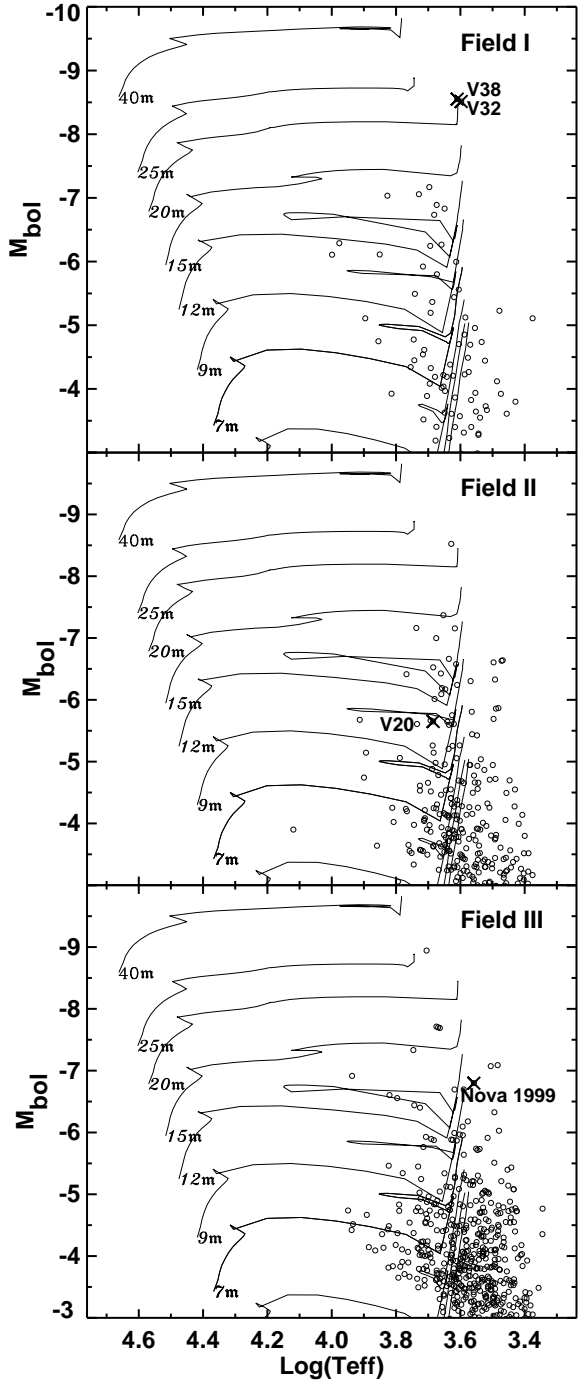}
\end{center}  
\caption{ H--R diagram for  IC\,1613.  The evolutionary tracks from
Charbonnel et al. (1993) for $Z=0.004$ are also shown. Known
variables are marked with dark crosses and labeled.
  \label{Fig10}}
\end{figure}

We needed to determine the effective temperatures ($T_{\rm eff}$)
and bolometric corrections (BC) in order to locate the stars on
the theoretical HRD. We used Costa \& Frogel's (1996) method to
transform K magnitudes and $(J-K)$ colors into $M_{\rm bol}$ and
$T_{\rm eff}$. We first transformed our $K$ magnitudes and
$(J-K)$ colors into the CIT system by means of the transformation
equations for the set of filters in use at "San Pedro Martir" observatory
(eq. 1 and 2 from Ruelas-Mayorga 1997). Then, using the method outlined by Costa \&
Frogel (1996), we transformed our $(J-K)_{CIT}$ colors to the
Johnson system and derived the bolometric correction for the $K$
magnitude, $M_{\rm bol}$  and $T_{\rm eff}$. We used the distance
modulus and reddening adopted above (Section~\ref{Sect6}). The
resulting HRDs are shown in Fig.~\ref{Fig10}. The evolutionary
tracks from Charbonnel et al. (1993) for $Z = 0.004$ were
superimposed on the same plot.

The brightest stars in Field I, V32 and V38, have bolometric
magnitudes of $M_{\rm bol} = -8.6 $ and masses of 20$\cal M_\odot$.
These values are very different from those found using $UBV$
photometry (Georgiev et al. 1999), which are larger by  $\approx
1\,$mag in $M_{\rm bol}$ and 20$\cal M_\odot$.  The largest sources of
uncertainty in transformation to the theoretical plane are the
uncertainties in $(J-K)$ colors.  Adopting a maximum uncertainty
in $(J-K)$ of $0.15\,$mag, Costa \& Frogel's (1996) equations
imply a maximum uncertainty of $0.25\,$mag in $M_{\rm bol}$ and
$400\,$K in $T_{\rm eff}$.  Therefore,
the differences in temperatures and bolometric magnitudes
determined from optical and infrared photometry are much larger
that the uncertainties due to the photometry or the transformation
to the theoretical plane, and confirm the well-known fact that it
is very unreliable to determine bolometric corrections for very
red stars from data at optical wavelengths. The effective
temperatures derived from our NIR data for V32 and V38 are
approximately 4000 K, in good agreement with their M0Ia spectral
class (Elias \& Frogel 1985).

The brightest star in Field II also has a mass of 20$\cal M_\odot$,
while the brightest star in Field III is more massive at 25$\cal M_\odot$.  
The fainter RSGs and brighter AGB stars in the
magnitude range $-5.5 <M_{\rm bol} < -7.5 $ have masses of 12 to
15$\cal M_\odot$. 

The only variable star in Field II is V20 --- a Cepheid with
$P=41\fd953$ and $B$ amplitude $\Delta_{B}=1.66$ mag (Sandage 1971).
The derived bolometric magnitude from our IR data is $M_{\rm bol}
= -5.7$, the effective temperature $\approx$ 5000 K and 9$\cal M_\odot$. 
Two WR/O candidates were also identified in Field II (W2
and W3: Armandroff \& Massey 1985, Azzopardi et al, 1988) but they 
are too faint for reliable measurements.

We have identified an object at the position of the recently
discovered Nova (King et al. 1999) in Field III. On the
our IR frames, taken on Oct. 15 1998, Nova 1999 has $(J-K)= 1.14$
and $K=14.69$.  If this is the same object, its
presence on our IR frames a year before its discovery  bring its 
classification as a Nova into question.
In the subsequent paper, we will discuss the light variations and
spectrum of this peculiar object.

\section{Distance to IC\,1613}

Using the apparent $J$ and $K$ magnitudes of the Cepheid variable
V20 and period-luminosity (PL) relation, we can obtain the
apparent distance moduli in J and K. Adopting Madore \& Freedman's
(1991) $J$ and $K$ calibrations of the PL relation, based upon 25
Cepheids in the LMC, and the period found by Sandage (1971), we
obtain distance moduli of $(m-M)_{J} = 24.42\pm0.2$ and $(m-M)_{K}
= 24.37\pm0.2$.  These are in satisfactory agreement with Cole et
al.'s (1999) determination of $(m-M)_{0} = 24.27\pm0.1$.

\section{The Bolometric luminosity function}

\begin{figure}
\begin{center}
\includegraphics[width=\columnwidth]{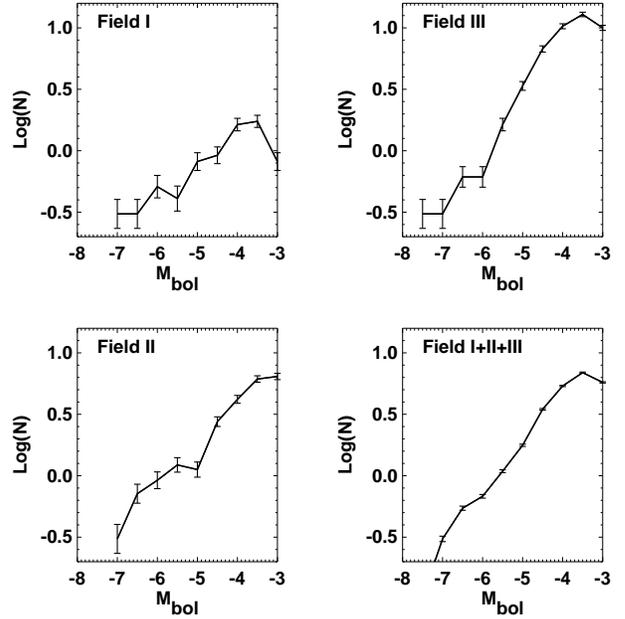}
\end{center}  
\caption{ The bolometric luminosity function for all stars in Field I,
Field II and Field III.  \lq\lq N" is the number of stars per 0.5
mag interval per square arcmin. The error bars represent the
uncertainties due to counting statistics.
  \label{Fig11}}
\end{figure}

The bolometric LFs (not corrected for completeness) for our three
fields in IC\,1613 are compared in  Fig.~\ref{Fig11}. All of the
LFs have a discontinuity at $-5.0 < M_{\rm bol} < -5.7 $, which
characterizes the boundaries between RSG and AGB stars.  In
Fig.~\ref{Fig11}, a LF formed from the sum of the stars in the
three fields is also shown. We would like to point out the lack of a
discontinuity in the LF for the AGB stars in all three fields.
The logarithmic LFs are linear for the AGB stars ($-3.0 <M_{\rm
bol} < -6.0 $), so they can be characterized by a power-law
exponent.  The derived by least squares fit exponent for the composite 
LF is 0.43$\pm0.05$.

\section{Summary}

In summary, we find substantial evolved stellar populations in our
three fields in IC\,1613, including the region of ongoing star
formation. The brightest stars are massive (20 - 25$\cal M_\odot$) red
supergiants with $K$ magnitudes between $13 - 13.5$ and $(J-K)
\approx 1$.  Based upon a comparison with the Padua isochrones
(Bertelli et al. 1994), these brightest RSGs are very young, with
ages between  8 and 25 Myr.

The main feature of the NIR CMDs is the rich intermediate age AGB
population. In the central part of the galaxy (our Fields II and
III), the tip of AGB stars can be located at $K = 16.20$. 
There is an indication of lack of ABG stars in Field I (centered on most active HII region
of the galaxy) that are redder than $(J-K)_{0} = 1$.  All of the
AGB stars in Field I are younger than 1 Gyr, whereas those in the
Fields II and III span ages from 1 to 10 Gyr. 
Using  the PL relation derived by Madore \& Freedman (1991) along
with our $J$ and $K$ magnitudes for the Cepheid variable V20 we
calculate a distance modulus $(m-M)_{K} = 24.37\pm0.2$. In Field
III we identify an object at the position of the recently
discovered Nova 1999 (King et al. 1999).  Its presence in our IR 
frames one year before discovery indicate that this star is too bright
 to be a classical Nova.

\acknowledgements

This work was performed while J.B. was a visiting astronomer in UNAM, 
Mexico under contacts CONACYT No. 27984E and DGAPA IN122298 and was supported
in part by the Bulgarian National Science Foundation grant under 
contract No. F-812/1998 with the Bulgarian Ministry of Education and Sciences.

 \end{document}